# Origin of Matching Effect in Anti-dot Array of Superconducting NbN Thin Films

Sanjeev Kumar[a], Chandan Kumar[a], John Jesudasan[b], Vivas Bagwe[b], Pradnya Parab[a], Pratap Raychaudhuri[b] and Sangita Bose[a1]

[a]UM-DAE Center for Excellence in Basic Sciences, University of Mumbai, Vidhyanagari Campus, Mumbai-400098, India.

[b]Tata Institute of Fundamental Research, Homi Bhabha Road, Colaba, Mumbai 400005.

We investigate the origin of matching effect observed in disordered superconducting NbN thin films with periodic array of holes. In addition to the periodic variation in the electrical resistance just above the superconducting transition temperature, $T_{c0}$, we find pronounced periodic variations with magnetic field in all dynamical quantities which can be influenced by flux-line motion under an external drive such as the magnetic shielding response and the critical current which survive in some samples down to temperatures as low as $0.09T_{c0}$. In contrast, the superconducting energy gap, $\Delta$, which is a true thermodynamic quantity does not show any periodic variation with magnetic fields for the same films. Our results show that commensurate pinning of the flux line lattice driven by vortex-vortex interaction is the dominant mechanism for the observed matching effects in these superconducting anti-dot films rather than Little-Parks like quantum interference effect.

---

[1]Email: sangita@cbs.ac.in

I.  **INTRODUCTION**

Matching effects which manifest itself as periodic oscillation in properties like magneto-resistance and critical currents with magnetic field has been traditionally observed in networks of thin walled superconducting cylinders or wire networks. The period of oscillation is proportional to a flux quantum ($\Phi_0$ =h/2e). In recent years, this effect has also been observed in superconducting films with periodic array of holes or anti-dots. In this geometry, the anti-dots provide a potential well to trap vortices or flux lines and has formed a subject of considerable theoretical [1,2,3,4,5,6,7,8,]and experimental attention [9,10,11,12,13,14,15,16,17,18,19,20,21,22]. Novel effects such as the formation of stable vortex-antivortex molecules in equilateral mesoscopic type I superconducting triangles have also been reported [23]. In addition there are predictions of "Rachet effects" or "multi-quanta states" in superconductors having random pinning density which can lead to the transport of interacting vortices [2]. Interestingly, studies have also been carried on disordered superconductors such as InO$_x$ where matching effects is observed to persist in the insulating side across the superconductor-insulator transition (SIT) obtained by tuning the disorder [24].

The origin of matching effects has been explained through two different, but not mutually exclusive mechanisms. The first one is the Little Parks like Quantum interference Effect (QI) seen in an array of superconducting loops where the super-current around each loop goes to zero at integral number of flux quantum. Here, the superconducting order parameter, and hence the superconducting transition temperature, $T_{c0}$, mimics the periodic variations in the supercurrents showing a maximum at the matching fields [25]. For QI to hold it is necessary that the width of the superconductor (in between the holes) should be comparable to the Ginzburg-Landau coherence length ($\xi_{GL}$). The second mechanism is related to vortex pinning. It is believed that the collective pinning of the vortex lattice is enhanced when each hole contains an integer number of vortices i.e. the vortex lattice becomes commensurate with the anti-dot lattice, thereby giving rise to a periodic variation in the critical current ($I_c$). This mechanism is referred to as commensurate pinning (CP). The essential difference between these two mechanisms is that while the former would result in periodic variations in all thermodynamic quantities, the latter would give periodic variations only in properties related to motion of flux lines. So far there is little consensus on the relative importance of these two effects [9,10,11,19]. This is primarily due to the fact that matching effects has been probed mainly through magneto-resistance (MR) oscillations at temperatures very close to $T_{c0}$, a quantity that can be affected both by oscillations in thermodynamic quantities such as $T_{c0}$ as well as the periodic variation in flux pinning. Recently, we developed a new method based on the magnetic shielding response of the sample, which allows the measurement of matching effect well below $T_{c0}$ [26]. Using this method we demonstrated that in anti-dot arrays of moderately disordered NbN films, the matching effect can persist down to temperatures as low as 0.09 $T_{c0}$.

In this paper, we address this issue by carrying out detailed investigation of the matching effect in different physical quantities, such as magneto-resistance, critical current, magnetic shielding response, transition temperature and the superconducting energy gap ($\Delta$),in anti-dot arrays on superconducting NbN thin films grown on anodic alumina membranes (AAM). Our results show that the matching effects are present in all the physical quantities which are associated with motion of the flux lines in the presence of an external drive. However, $\Delta$, which is a true thermodynamic quantity does not show any matching effect in the same temperature range. These observations clearly demonstrate that the matching effect in anti-dot arrays is primarily governed by commensurate trapping of flux in the periodic holes of the anti-dot array of the superconducting films.

## II. EXPERIMENTAL DETAILS

Our samples consist of superconducting NbN films deposited by reactive DC magnetron sputtering on free-standing nanoporous anodic alumina membranes (AAM) obtained from Synkera technologies. The thin films were formed by sputtering Nb in $Ar/N_2$ gas mixture keeping the substrate at $600^0$ C. The AAM had a pore diameter of 18 nm with an inter-pore separation of 44 nm. To check for consistency, some measurements were also carried out on films grown on AAM with 35 nm pore diameter and inter-pore separation of 95 nm [27]. On both these AAMs, films with different levels of disorder were grown. The disorder in these films is controlled by controlling the Nb vacancies in NbN by tuning the $Ar/N_2$ ratio. For each sample, the superconducting transition was measured using both electrical resistivity and diamagnetic shielding response. We define the transition temperature as the temperature where the global zero resistance state is established and will denote it by $T_{c0}$. Operationally this corresponds to the temperature where the resistance goes below our measurable limit or where we observe the onset of the diamagnetic shielding response. The least disordered samples studied has a $T_{c0} \sim 12.1$ K and the most disordered films had a $T_{c0} \sim 4.2$ K. The samples are labelled as S-x-y, where x stands for the pore diameter and y stands for the $T_{c0}$ value rounded to the nearest integer. These NbN films were patterned in different geometries for different measurements e.g., circular samples with 8mm diameter for measuring the shielding response using the mutual inductance set-up and thin striplines of dimensions 7 mm x 3 mm for critical current measurements. The magnetic shielding response was measured using two-coil mutual inductance technique operating at 60 kHz. In this technique the circular film is sandwiched between a quadrupolar primary coil and a dipolar secondary coil [26] and the magnetic shielding response of the sample is measured through the mutual inductance between the primary and the secondary. The ac excitation field of the primary is fixed to a peak-to-peak value of 10 mOe field, where the shielding response is in the linear regime. All transport measurements were performed using the standard four probe technique. The measurements were performed either in a conventional $^4$He flow cryostat fitted with an 8 T superconducting magnet down to 2 K or in a $^3$He cryostat fitted with a 6 T superconducting magnet down to 300 mK.

In order to measure the superconducting energy gap planar tunnel junctions of NbN/Nb$_2$O$_5$/Ag were fabricated in the following way: a NbN thin film of width 300μm was first deposited on alumina templates which were then oxidized at 200°C for two hours in ambient conditions. Counter electrodes of Ag was evaporated in perpendicular geometry to NbN film which were of similar width as that of the NbN films. Provision was made to measure the current versus voltage (I-V) characteristics across the tunnel junction and the longitudinal resistivity versus temperature (R-T) of the NbN film on the same device (schematic of the device is shown in Figure 5). The differential conductance of the tunnel junctions $G(V) = \left.\frac{dI}{dV}\right|_V$ was obtained by numerically differentiating the I-V curves.

**RESULTS**

**A. Matching effect in magneto-resistance, magnetic shielding response and critical current**

Figure 1(b)-(d) show the resistance as a function of magnetic field at different temperatures in the three samples S_18_12, S_18_6 and S_18_4 respectively. All three samples show oscillations in magneto-resistance (MR), in the form of sharp minima in the resistance

below the onset of the superconducting transition, $T_c^{on}$ (shown in Fig. 1(a)) which gradually disappear as the sample is cooled below $T_{c0}$. The value of the matching field ($B_M$) is consistent with the theoretically expected value of $B_M = n\Phi_0/A$ where $A = 2\Phi_0/\sqrt{3}d^2$. The matching effect is observed upto two matching fields. Similar measurements on the films grown on AAM with 35 nm pore diameter show the matching effect up to 4 matching fields [27].

Figure 2 (a) - (c) show the magnetic field dependence of the real part of the mutual inductance M', for three of the NbN circular films at various temperatures in the superconducting state. All the films show an oscillatory behavior with pronounced minima at the matching fields. The oscillations in mutual inductance persist down to much lower temperatures compared to MR oscillations. For the sample S_18_12 the oscillations persists down to 6 K which is ~ $0.5T_{c0}$. The temperature window over which oscillations persists goes on increasing with increase in disorder. For the most disordered sample, S_18_4 the oscillations survive down to 300 mK which corresponds to $0.09T_{c0}$.

As a further consistency check we also measured the magnetic field variation of $I_c$ in similar films grown in the stripline geometry. $I_c$ is extracted from the current voltage characteristics and $I_c$ is taken as the current at which the voltage appears. The magnetic field dependence of the critical current $I_c$ at T = 5 K for the sample S_18_9 is shown in figure 3(b). (The inset of the figure shows the R-T for the same film used to determine the $T_{c0}$). The matching effect in $I_c$ is manifested as pronounced maxima at the same matching fields which is consistent with the minima observed in mutual inductance (M′) and resistance (R) measurements (See figure 3(a)). It may be possible that the oscillations persist down to even lower temperatures, which we could not measure as heating effects became appreciable due to increase in critical currents.

## B. Matching effect is the superconducting transition temperature, $T_c$

We now investigate the matching effect in $T_{c0}$. We look at two quantities: The onset temperature, $T_c^{on}$, where the resistance reaches 90% of the normal state value, and $T_{c0}$, where the resistance goes below our measurable limit (See figure 1(a)). The latter coincides with the onset of the diamagnetic response, which was independently verified from the magnetic shielding response measurements. In the absence of magnetic field, both these quantities are related to the formation of Cooper pairs, while the difference between $T_{c0}$ and $T_c^{on}$ can be primarily ascribed to inhomogeneity in the sample and a very small temperature window of Ginzburg-Landau fluctuations. However, in the presence of magnetic field the two temperatures signify two different physical processes. $T_c^{on}$ is determined by the temperature where the system undergoes transition from the normal state to the mixed state. However, the resistance continues to remain non-zero down to lower temperature, $T_{c0}$, till the flux flow under combined influence of thermal activation and external drive current is arrested by the pinning potential. Thus $T_{c0}$ denotes the onset of flux flow, whereas $T_c^{on}$ continues to be determined by the thermodynamic superconducting transition. In order to see the variation of $T_{c0}$ and $T_c^{on}$ with magnetic field, we measured the resistance as a function of temperature for the films S_18_4, S_18_6 and S_18_12 at different magnetic fields (Fig. 4 (a)-(c)). Figures 4(d)-(f) show the magnetic field variation of $T_{c0}$ and $T_c^{on}$ extracted from these data. $T_{c0}$ shows pronounced oscillations with magnetic field (with a maximum at each matching field) suggesting a periodic variation in the flux line pinning strength. The amplitude of $T_{c0}$ oscillation is ~ 500 mK for all the three samples. On the other hand, $T_c^{on}$ decreases monotonically by about 1.6% with magnetic field with no signature of matching effects.

**C. Measurement of the superconducting energy gap for anti-dot array**

To explore if the matching effect is observed in a true thermodynamic quantity we measure the magnetic field variation of the superconducting energy gap. Figure 5 (b) shows the differential conductance $G(V) = dI/dV|_V$ at various temperatures down to $T = 2\ K$ for a tunnel junction fabricated on sample S_18_8. At each temperature the maximum bias voltage is limited by the current reaching the critical current of the superconductor. All the $G(V)$ vs. $V$ spectra were fitted with the tunneling equation,

$$G(V) = \frac{dI}{dV} = \frac{d}{dV}[\frac{1}{R_N}\int_{-\infty}^{+\infty} N_S(E)N_N(E-eV)(f(E)-f(E-eV)\}dE]$$

where $N_S(E)$ and $N_N(E)$ are the normalized density of states for the superconducting and normal metal respectively, $f(E)$ is the Fermi Dirac distribution function and $R_N$ being the resistance of the tunnel junction for $V \gg \Delta/e$. $N_S(E)$ is given by $N_S(E) = Re\left\{\frac{E-i\Gamma}{\left([(E-i\Gamma)^2-\Delta^2]^{\frac{1}{2}}\right)}\right\}$ where $\Delta$ is the superconducting energy gap and $\Gamma$ is the phenomenological broadening parameter. We observe that above theoretical expression for tunneling conductance fits very well to the experimental data for our sample in the whole temperature range. We investigate the possibility of matching effects in $\Delta$, by measuring the $G(V)$ vs. $V$ spectra at the matching fields and midway between two matching fields at different temperatures from 1.8 K to 5.5 K. The latter corresponds to a reduced temperature, $t = T/T_{c0} \sim 0.70$, at which we observe pronounced matching effect in the shielding response in all the samples reported in this paper. At 1.8 K the conductance spectra at different field lie over each other with no detectable difference (Fig. 5(c)). At 5.5 K, where the spectra are restricted to bias voltage below 0.9 mV (due to onset of critical current) we concentrate on the zero bias conductance (ZBC) (Fig. 5(d)). The ZBC is constant with magnetic field without any signature of matching effect. We further confirmed that the same tunnel junction sample showed matching effects in magneto-resistance at temperatures close to $T_{c0}$ (shown in Fig. 5(a)).

Since the absence of matching effects in $\Delta$, is a strong evidence of the absence of QI as the dominant mechanism for matching effects, we analyze the data more critically. We first focus on the magnetic field variation of $\Delta$ at low temperature ( 2.0 K ). Within BCS theory, $\frac{\Delta(T \to 0)}{k_B T_c^M}$ (where $T_c^M$ is the mean field transition temperature) is a constant determined by the value of the attractive pairing interaction. Thus if QI dominates, the variation of $T_{c0}$ observed in our experiment should reflect the variation in $T_c^M$, and therefore a proportional change in $\Delta$. Since with magnetic field $\Delta T_c \sim 500$ mK, we would expect a corresponding 5% variation in $\Delta(T \to 0)$ with magnetic field arising from QI effect. To check whether this small variation is within the resolution limit of our measurements, in Fig. 6(a) we plot the G(V) vs V spectra at 2.0 K along with the theoretical BCS curves calculated by varying the $\Delta$ from its best fit value by 5% (keeping $\Gamma$ constant). These curves are clearly outside the noise limit of our measurement, showing that our measurement would have picked up the variation if it was present. The similar plots at 4.0 and 5.5 K (Fig. 6(b) and (c)) clearly shows that at these temperature the signature of 5% change in $\Delta$, would have reflected in the magnetic field variation in ZBC beyond the noise level. This confirms that the matching effect in $T_{c0}$ results from CP rather than QI as discussed Section B. On the other hand, $T_c^{on}$ shows a very small variation with $H$, consistent with the nearly constant value of $\Delta$.

**IV. DISCUSSIONS**

The emerging picture from our measurements is that the matching effect clearly manifest only in driven quantities such as critical current, $T_{c0}$ and a.c. magnetic shielding response, whereas thermodynamic quantities such as $\Delta$ or $T_c^{on}$ do not show any periodic variation with magnetic field. This is a clear indicator that the matching effect results from periodic variation of the vortex pinning strength rather than a periodic variation of the amplitude of the superconducting order parameter (such as in the classic Little-Parks experiment) caused by QI, which reflects in all measurable quantities.

We now analyze whether CP is consistent with other aspects of our data. The origin of commensurate pinning of the flux tubes in an anti-dot lattice is due to the competition of two restoring forces arising from the confining potential created by the surrounding superconductor and the repulsive interaction potential between adjacent pearl vortices trapped in the anti-dots. Interestingly, the confining potential will individually pin the vortices inside the holes. However, the restoring force from the inter-vortex repulsive interaction will increase the pinning at the matching fields by confining the flux lines in a "cage" formed by the surrounding flux tubes. At the matching fields, the enhancement of pining should lead to maxima in critical current while the magneto-resistance and shielding response should show minima, which is consistent with our observations. Considering the repulsive interaction between adjacent flux tubes in an array, the maximum number of multiquanta vortices that can get accommodated in a given hole of radius R is given by $n_s = [R/2\xi(T)]^2$. This was worked out by Doria *et. al.* considering an array of vortices with inter-vortex interactions [28]. This number for the NbN thin films with the periodic array of holes turns out to be ~2 for the film with the anti-dot diameter of 18 nm and ~4 for the film with the anti-dot diameter of 35 nm at temperatures $0.85T_{c0}$ (where, $\xi(0)$ = 6 nm, as estimated previously [29]) which is consistent with our observations. Very close to $T_{c0}$, this number should decrease as $\xi(T)$ increases which was observed in the magnetoresistance measurements. This indicates that inter-vortex interactions need to be considered in the interpretation of our results.

Inter-vortex interactions can also explain one of the distinct results of our experiments i.e. the survival of the matching effects deep in the superconducting state and the increase of the temperature window for their observation with increase in disorder of the films *(~ 0.5$T_{c0}$ for the sample S_18_12 and ~ 0.09$T_{c0}$ for the sample S_18_4)*. This has been discussed in detail in Ref [26] where we have reasoned that the presence of defects and in-homogeneities in the most disordered films reduces the confining potential barrier for the trapping of vortices. In addition, the inter-vortex repulsive interaction also increases with disorder as the penetration depth changes by almost an order of magnitude in these films (as shown in Ref. 30). Hence, for films with the same diameter of the anti-dot, *a*, the film with more disorder should show more pronounced matching effects which is consistent with our observations.

We now discuss the matching effect observed in $T_{c0}$. As discussed previously, $T_{c0}$ in our experiments represents the temperature at which the motion of the trapped flux sets in. At the matching fields, the pinning is strongest. Therefore, the vortices find it difficult to overcome the thermal activation barrier and the flux motion starts at a higher temperature compared to that at non-matching fields where the pinning potential is shallow and the vortices easily overcome the thermal activation due to pinning, resulting in lower $T_{c0}$. This leads to broader transitions at the non-matching fields as the mean field transition temperature (or $T_c^{on}$) is almost independent of the magnetic field (in this field range). Hence, CP can explain the oscillations observed in $T_{c0}$ in our films.

In conclusion, we have investigated the origin of matching effect in disordered superconducting NbN thin films with periodic array of holes. We have measured different physical quantities like the magneto-resistance, critical current, dynamical screening response, critical temperature and the superconducting energy gap. In our experiments, all dynamical quantities which can be influenced by the flux line motion under an external drive showed pronounced matching effects. However, the superconducting energy gap which is a true thermodynamic quantity did not show any periodic variation with magnetic fields for the same films. In addition the temperature window for the survival of the matching effect increased with increase in the disorder of the films and it extended to as low as *$0.09T_{c0}$* for the most disordered film. Our results indicate that CP leading to vortex-vortex interaction is the dominant mechanism for the observed matching effects in these superconducting anti-dot films.

**Figure Captions**

**Figure 1: (Colour Online)**

(a) Temperature variation of resistance for the films S_18_12, S_18_6 and S_18_4 showing the transition temperature, $T_{c0}$ and the onset temperature $T_c^{on}$. The resistance has been normalized by the value in the normal state (i.e. by $R_N$ at T = 14.5 K for all the films). $T_c^{on}$ is the temperature at which resistance drops to 90% of the normal state value while the $T_{c0}$ is the temperature where the resistance disappears.

(b)–(d) Variation of resistance with magnetic field (B) at different temperatures below $T_c^{on}$ for the films S_18_12, S_18_6 and S_18_4 respectively. For S_18_12 in (b) the plots are shown for temperatures 11.0K, 11.5K, 12.0K, 12.5K and 13.0K. For S_18_6 in (c) the plots are shown for temperatures 6.5K, 7.0K, 7.5K and 8.0K. For S_18_4 in (d) the plots are shown for temperatures 3.0K, 3.5K, 3.9K, 4.2K and 4.5K.

**Figure 2: (Colour Online)**

(a)-(c) Magnetic field (B) variation of the real part of the mutual inductance (M′) at different temperatures below $T_{c0}$ for the films S_18_12, S_18_6 and S_18_4 respectively. For S_18_12 in (a) the plots are shown for temperatures 6.0K, 8.0K, 9.5K, 10.5K and 11.5K. For S_18_6 in (b) the plots are shown for temperatures 1.7K, 3.5K, 4.5K, 5.0K and 5.5K. For S_18_4 in (c) the plots are shown for temperatures 0.38K, 1.5K, 2.5K, 3.0K and 3.5K.

**Figure 3: (Colour Online)**

Variation of (a) resistance (R) at temperature T = 8.25K and (b) critical current ($I_c$) at temperature T = 5K for the sample S_18_9 with magnetic field (B). The resistance versus temperature (R vs T) for the same sample is shown in the inset in (b). The magnetic field variation of resistance was done with a current of 20μamp which is much lower than the critical current of the sample at this temperature.

**Figure 4: (Colour Online)**

(a)-(c) Temperature variation of resistance (R) at various matching and non-matching fields for the films S_18_12, S_18_6 and S_18_4 respectively. The plots for all three samples are shown for the magnetic fields indicated in Figure 4(b).

(d)-(f) Variation of $T_c^{on}$ and $T_{c0}$ with magnetic field normalized with respect to the matching field ($B_M$) for the films S_18_12, S_18_6 and S_18_4 respectively.

**Figure 5: (Colour Online)**

(a) Variation of resistance (R) with magnetic field (B) at T = 7.4K (below $T_{c0}$) for the film S_18_8. On the same sample the tunnel junction was fabricated on which the differential conductance measurements have been carried out. The schematic of the device is shown at the bottom of the figure.

(b) Differential conductance (dI/dV) as a function of bias voltage (V) at various temperatures down to T = 2 K for the film, S_18_8. Open circles represent data while the solid lines are the theoretical fits using the tunneling equation (given in the text). The plots are shown for the temperatures 2.0K, 3.0K, 3.5K, 4.0K, 4.5K, 5.0K and 5.5K.

(c) Differential conductance as a function of bias voltage (V) for the same sample at various matching and non-matching fields at T = 1.8 K. The magnetic field was varied from 0 to 18kG in steps of 4.5kG. The Δ value (obtained from the fits using the tunneling equation) is plotted as a function of magnetic field (B) in the same graph. The scale for the same is shown at the right and top respectively.

(d) Differential conductance as a function of bias voltage (V) for the same sample at various matching and non-matching fields at T = 5.5 K. The magnetic field was varied from 0 to 18kG in steps of 4.5kG. The zero bias conductance value, ZBC is plotted as a function of magnetic field (B) in the same graph. The scale for the same is shown at the right and top respectively.

**Figure 6: (Colour Online)**

(a)-(c) Differential conductance (dI/dV) as a function of bias voltage (V) at temperatures T = 2.0 K, T = 4.0 K and T = 5.5 K for the sample S_18_8. Open circles represent experimental data while the lines are the simulated curves using the tunneling equation for a fixed Γ and different Δ. The solid black line is the best fit to the data. The other two curves (for each temperature) are obtained by keeping the Γ same and changing Δ by 5% about the Δ value which best fits the experimental data.

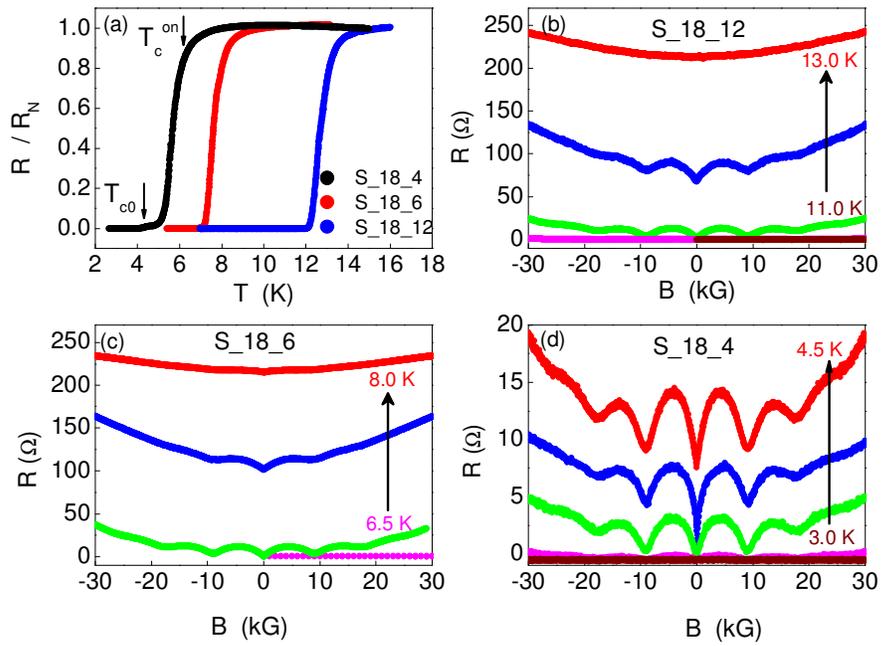

Figure 1

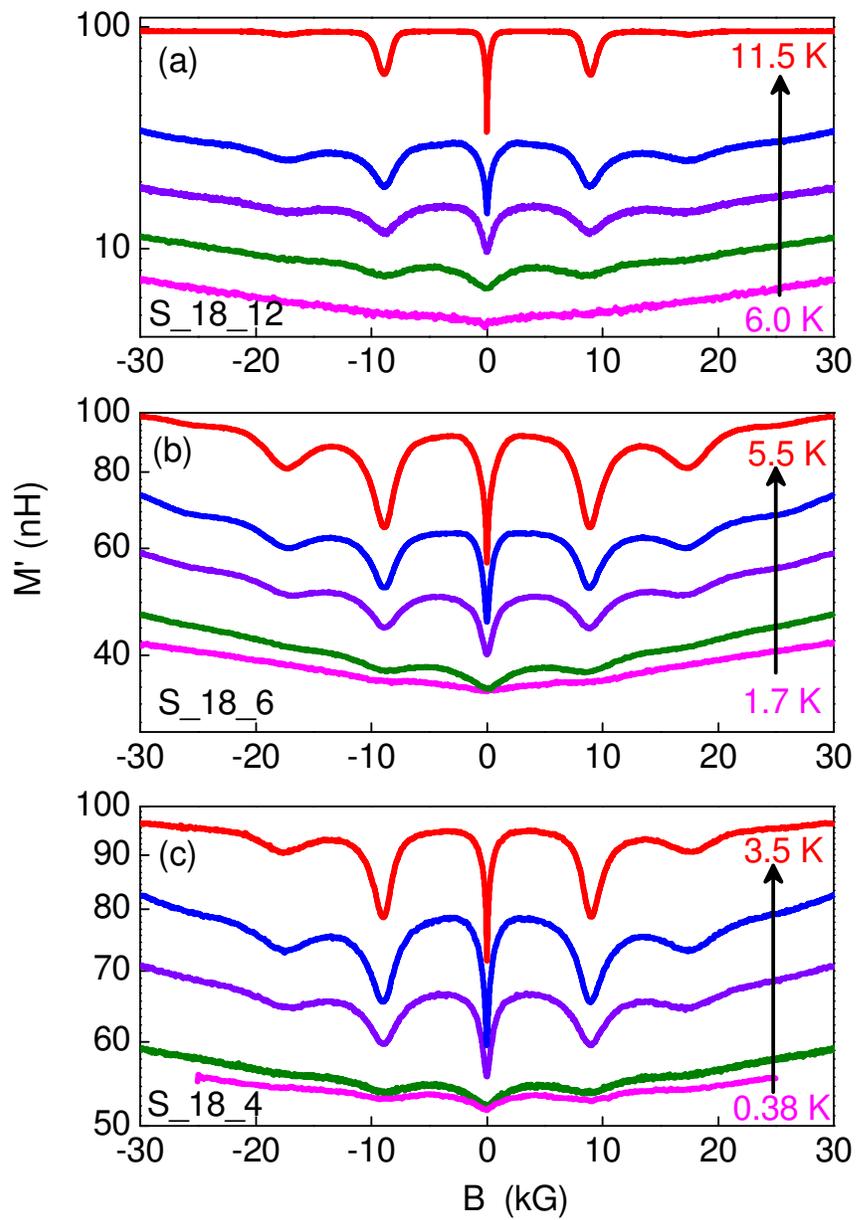

Figure 2

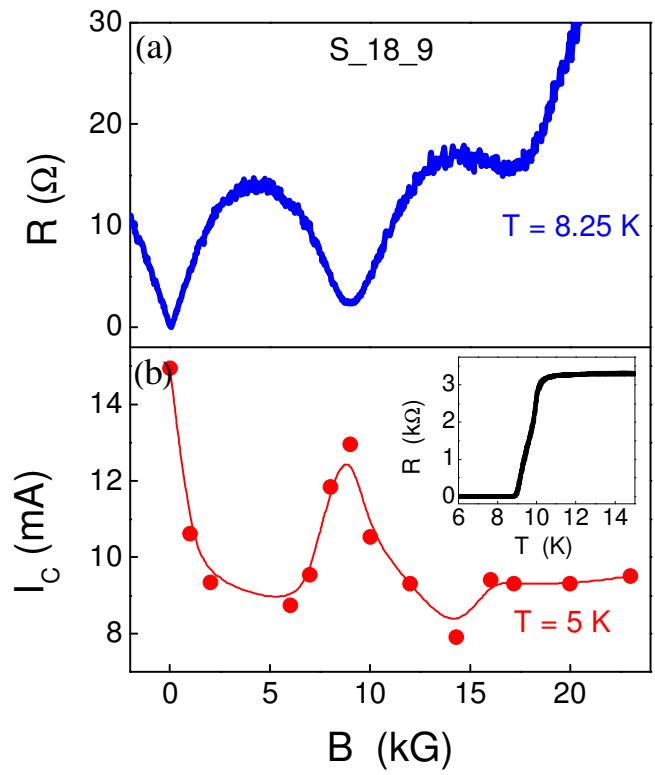

Figure 3

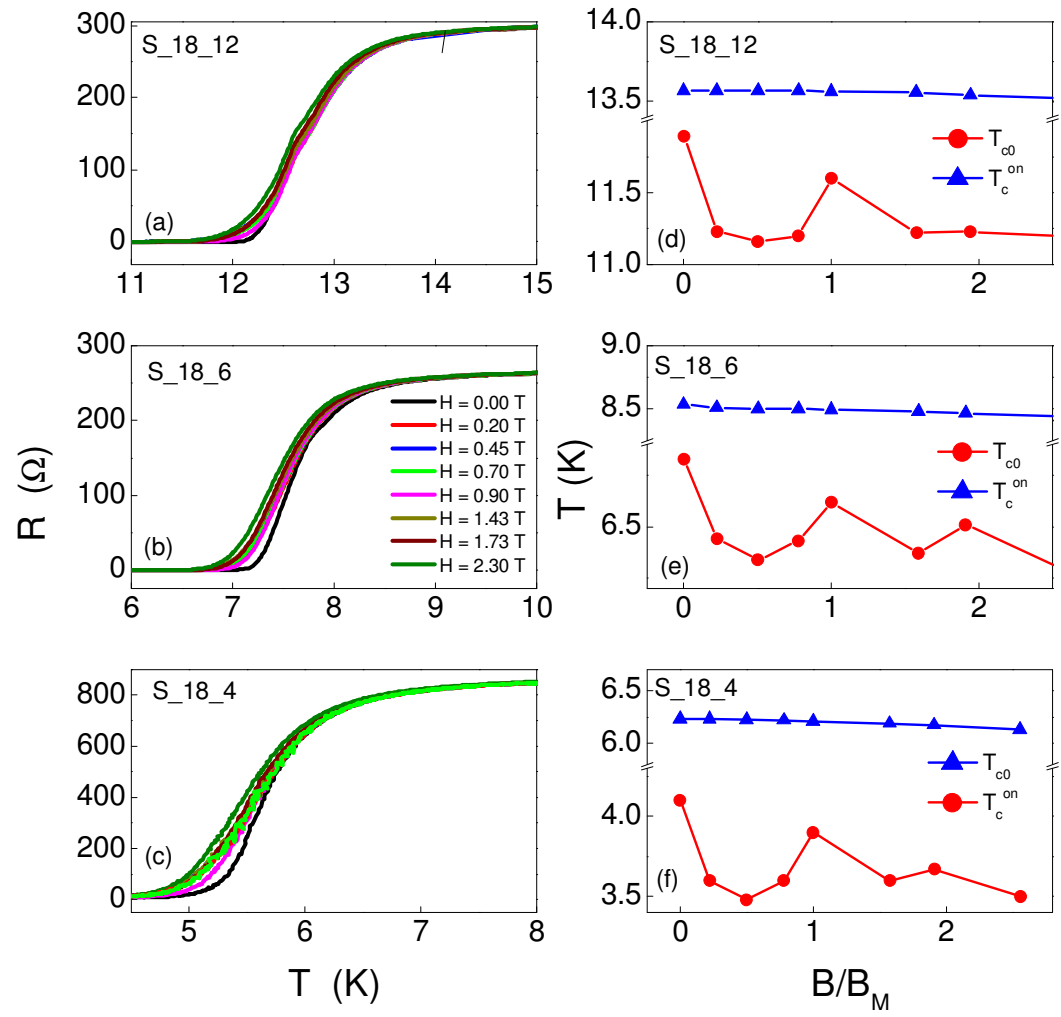

Figure 4

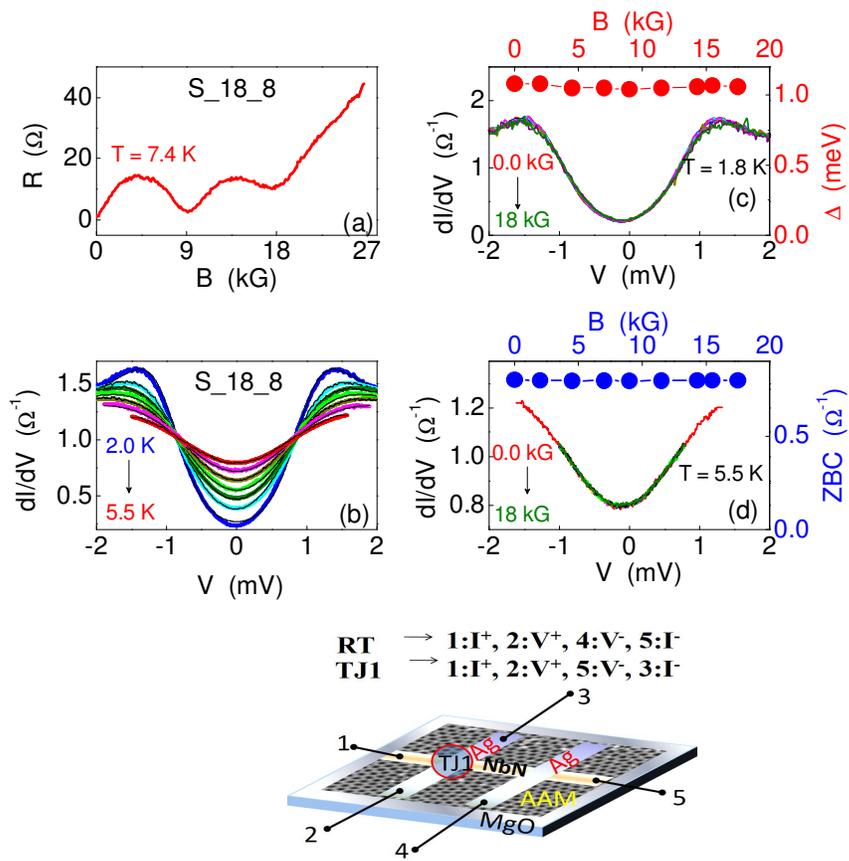

Figure 5

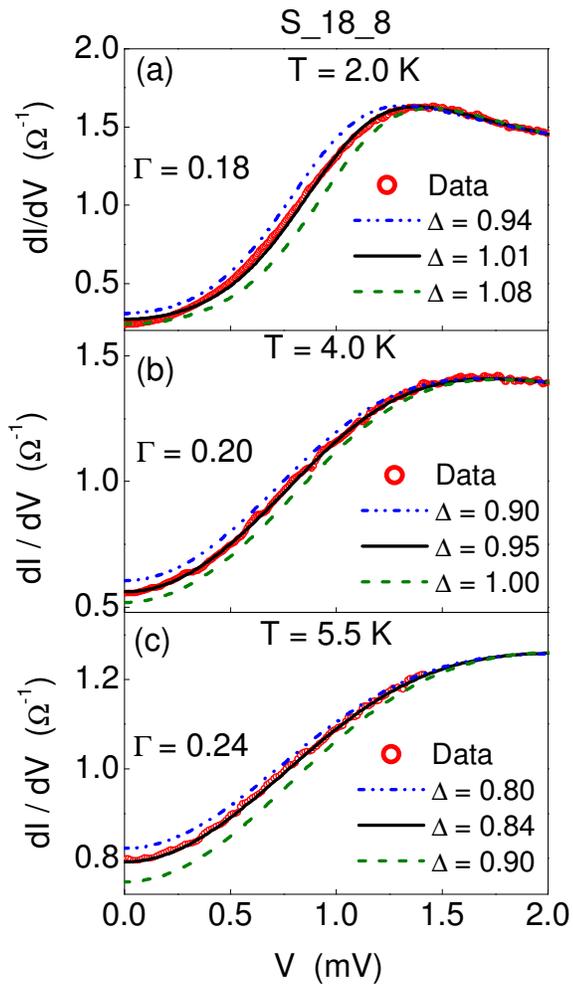

Figure 6